\documentstyle[preprint,eqsecnum,aps]{revtex}
\input boxedeps.tex
\SetRokickiEPSFSpecial  
\HideDisplacementBoxes

\newcommand{\D}{\overline{D}}
\newcommand{\da}{\dagger}  
\newcommand{\be}{\begin{equation}}
\newcommand{\eq}{\end{equation}}
\newcommand{\Tr}{{\rm \, Tr \!}}    
\newcommand{\dm}{{\cal M}}          
\newcommand{\newl}{l}               

\begin{document}
\preprint{\parbox{1.5in}{DAMTP-2000-119}}
\title{Precision study of large-$N$ Yang-Mills theory\\
in $2+1$ dimensions}

\author{Simon Dalley}
\address{Centre for Mathematical Sciences, Cambridge University \\
Wilberforce Road, Cambridge CB3 0WA, England}

\author{Brett van de Sande}
\address{Geneva College,\\
3200 College Ave., Beaver Falls, PA~~15010}


\maketitle

\begin{abstract}
The boundstate problem in 2+1-dimensional large-$N$ Yang-Mills theory 
is accurately solved using the light-front Hamiltonian of transverse lattice
gauge theory. We conduct a thorough investigation of the space of
couplings on coarse lattices, 
finding a single renormalised trajectory on which Poincar\'e
symmetries are enhanced in boundstate solutions.
Augmented by existing data from finite-$N$ Euclidean lattice
simulations, we obtain accurate estimates of the
low-lying glueball spectrum at $N=\infty$.
\end{abstract}

\pacs{
{\tt$\backslash$\string pacs\{\}} should always be input,
even if empty.}
\narrowtext

\section{Introduction}
Yang-Mills theories are theoretically interesting
in $2+1$ dimensions because their properties are very similar to the
corresponding theory in $3+1$ dimensions, yet they 
can be handled much more accurately; see Ref.~\cite{teper1} for
a review of properties and  extensive references. 
They appear to exhibit linear confinement of heavy
sources, 
a discrete spectrum of (glueball) boundstates, and a
finite-temperature transition. 
Teper has recently performed a  comprehensive analysis,
using the standard tools of Euclidean $SU(N)$ lattice gauge theory,
of 2+1-dimensional Yang-Mills theories at $N = 2$, 3, 4,  and 5.
Hamiltonian lattice calculations have also recently been performed for
finite $N$ \cite{ham1,ham2}
and, though less comprehensive, the results are mainly consistent. 
With data at enough values of $N$,
one can contemplate an extrapolation to $N = \infty$.
This is a limit of special interest for 
`analytic' approaches to gauge theory, 
which often take advantage of large-$N$ simplifications.
In the absence of any other criteria for the errors involved, 
the only way to know how well `analytic' approaches
are doing, for example those of Refs.~\cite{nair,gravity},
is to compare with lattice data and their extrapolation to large
$N$.

A related question is: how close is $N= \infty$ to small $N$?
This question can only be faithfully answered once there are
accurate results in both limits. The $1/N$ expansion \cite{hoof}
is typically an asymptotic one and, {\em a priori},
observables in the two limits need not
be close in value. The existing finite-$N$ data suggest 
strong 
suppression of corrections to the large-$N$ limit \cite{teper1,teper2}, a
conclusion that was speculated about much earlier \cite{marek}, 
on the basis of less reliable lattice data. If true, this fact
deserves a deeper understanding.

The main objective of this paper
is to address these issues for $2+1$ dimensional Yang-Mills theory
with explicit calculations at $N= \infty$.
In Refs.~\cite{us1,us2} (see also \cite{burk})
 we used the large-$N$ limit of $2+1$-dimensional
Yang-Mills as a test for developing the transverse lattice 
method of solving non-abelian gauge theories \cite{bard}. 
Based on an
improved understanding of the sources of error in that 
calculation, we perform here a calculation at the next level
of approximation. We obtain a renormalised light-front Hamiltonian
on the transverse lattice for both the pure-glue and heavy-source
sector. From this, we obtain the glueball spectrum and the heavy-source
potential.
Existing finite-$N$ data, combined with our explicit large-$N$
results, 
are
used to determined the first few coefficients of the $1/N^{2}$
expansion of glueball masses in string tension units.

In the next section we briefly review the transverse lattice method,
the details of which have been covered elsewhere \cite{us1,us2,us3}. Section
3 describes the numerical search for the renormalised Hamiltonian
via tests of Poincar\'e invariance. 
Our thorough investigation yields a single, well-defined 
candidate for the renormalised trajectory in coupling space.
Results for the low-lying
glueball eigenstates on this trajectory and the first few coefficients of 
the $1/N^{2}$ expansion of their masses
are given. This improves upon current estimates of the
large-$N$ limit, allowing us to accurately verify that corrections
to it are highly suppressed.
In the conclusions we discuss possible reasons 
for the success of analytic approaches,
given their approximations.

\section{Transverse Lattice in $2+1$ Dimensions}

Adapted to $2+1$ Yang-Mills theory, the Bardeen-Pearson
transverse lattice gauge theory 
consists of continuum gauge potentials $\{A_{0},A_{2}\}$ and
space-time
co-ordinates $\{x^0,x^2\}$, together with gauge-covariant 
transverse link variables $M(x^1)$ running between
sites at $x^1$ and $x^1 + a$ on a transverse lattice of spacing $a$.
We also use the light-front combinations $x^{\pm} = (x^0 \pm x^2)/\sqrt{2}$,
$A^{\pm} = (A^0 \pm A^2)/\sqrt{2}$, {\em et cetera}. 
DLCQ \cite{dlcq} and Tamm-Dancoff  
cut-offs on the number of partons are used as intermediate 
regulators. These are extrapolated following the analysis of
ref.\cite{us2}. DLCQ means that we impose
anti-periodic boundary conditions  $x^- \sim x^- + 2\pi K/ P^+$, 
where $P^+$ is the total
light-front momentum, and $K$ is an integer cut-off. 
$x^+$ remains continuous and infinite, and is 
used as a canonical time variable to derive a
light-front Hamiltonian $P^-$. 
The most general action, from which $P^-$ will be derived canonically,
must allow all
gauge invariant operators that respect the 
Poincar\'e symmetries unviolated by the
(gauge invariant) cut-offs. 
Since we will explicitly extrapolate the DLCQ and Tamm-Dancoff
cut-offs, 
only local dimension 2 operators with respect to $\{x^+, x^-\}$
co-ordinates will be included at the outset. 
In the discussion hereafter we assume
this limit has be taken.

Near the transverse continuum limit $a \to 0$ corresponding to 
$SU(N)$ Yang-Mills, one expects $M$ to
take values in $SU(N)$. However, away from this limit
one can allow $M$ to be a general $N$x$N$ complex matrix, provided it still
gauge transforms covariantly. One must then search this
larger class of lattice theories for the renormalised 
trajectory that leads one
to the continuum limit $SU(N)$ Yang-Mills theory. 
Physical results are invariant along this trajectory and equal to
the values in the full continuum limit. The trajectory may be found by
renormalisation group transformations in the neighborhood of a fixed
point (continuum limit). However, this is difficult for the present 
formulation.
There are (roughly)
two possibilities for the behaviour of $M$ at a given point in the
space of couplings constants:
$M$ is a massive degree of freedom ($M=0$ is the classical minimum); or,
the `radial' part of $M$ condenses. We expect the latter to be the case
near the $a = 0$ limit of Yang-Mills, 
where the action should be minimized near values of $M$
in $SU(N)$ rather than $M=0$. 
Dealing with the condensation of the radial part, or using unitary
matrices for $M$ from the outset \cite{griffin}, is tricky in light-front
quantisation. This is what makes an analysis near $a=0$ difficult.
On the other hand, it is straightforward to perform 
canonical light-front quantisation about $M=0$, when this is a
stable minimum. If the renormalised
trajectory passes into such a region, we can then study it.

An alternative way to find the renormalised trajectory is to use 
symmetry \cite{perry}.
Generally speaking, we can define a quantum field theory by 
symmetry --- in our case gauge and Poincar\'e invariance ---
and a particular continuum limit (there may be more than one).
There is actually no reason why we cannot take a
partial continuum limit, a limit in some space-time directions but not in 
others, since Poincar\'e invariance should relate them. 
Thus, in Ref.~\cite{us2} we proposed to take the continuum limit of
Yang-Mills theory in the $\{x^0,x^2\}$ directions, and
tune couplings to impose full Poincar\'e invariance at finite 
transverse cut-off $a$.
This procedure can be carried out using light-front quantisation about $M=0$. 
Although this regime apparently cannot contain the 
Poincar\'e-invariant theory at $a=0$,
numerical evidence for the existence of a renormalised trajectory
was given, and has been extended to $3+1$
Yang-Mills \cite{us3}.  
In this paper, we present 
conclusive numerical evidence for the case of $2+1$ Yang-Mills.

For practical calculations, the remaining allowed operators in the action
must be pared down to a finite number of independent parameters, and
one must find some reasonable criteria to test Poincar\'e invariance.
We now develop these necessary approximations, following closely
our previous work.

\subsection{Pure-glue sector}

To reduce the space of couplings to a finite dimension, we use
various approximations:
\begin{enumerate}
\item quadratic canonical momentum operator $P^+$,
\item light-front momentum-independent couplings,
\item transverse locality, and
\item expansion in gauge-invariant powers of $M$. \label{expansion}
\end{enumerate}
The reasoning behind them is described in more detail in Ref.~\cite{us3}.
We only note here, that a poor choice of approximation
will simply mean that we cannot get close to the renormalised
trajectory, if it exists, and accuracy will suffer accordingly.
The principle physical approximation, Item~\ref{expansion},
is the `colour-dielectric'
expansion about $M=0$, which is applied to the light-cone gauge-fixed
Hamiltonian rather than the action.

We have studied the light-cone Hamiltonian derived from the
following $SU(N)$ gauge-invariant
action in the large-$N$ limit 
\begin{equation} 
A = \int dx^0\, dx^2\, \sum_{{x^1}} 
\D_{\alpha} M({x^1}) (\D^{\alpha} M({x^1}))^\da 
- V_{x^1} -{1 \over 2 G^2} \Tr \left\{ F^{\alpha \beta} F_{\alpha
\beta} \right\}
\end{equation}
where $\alpha \in \{ 0,2\}$ and 
\be
        \D_{\alpha} M({ x^1})  
        =  \left(\partial_{\alpha} +i A_{\alpha} ({ x^1})\right)
        M({ x^1})  
        -  i M({ x^1})   A_{\alpha}({{x^1}+a}) \;,
\label{covdiv}
\eq
is the tranvserse lattice covariant derivative. The `potential' term is
\begin{eqnarray}
 V_{x^1} & = &
\mu^2  \Tr\left\{M({ x^1})M^{\da}({ x^1})\right\} 
         + {\lambda_1 \over a N}
        \Tr\left\{ M ({x^1})M^{\da}({ x^1})M({x^1})M^{\da}({x^1}) 
\right\} \nonumber \\ && + {\lambda_2 \over a N}  
        \Tr\left\{ M({ x^1})M({ x^1}+a)M^{\da}({ x^1}+a)
         M^{\da}({ x^1}) 
\right\}  +  {\lambda_3 \over a N^2} 
        \left( \Tr\left\{ M({ x^1})M^{\da}({ x^1}) \right\}
         \right)^2 \; .
\end{eqnarray}
In light-cone gauge $A_{-}=0$ and after eliminating $A_{+}$ by its
(constraint) equation of motion, the corresponding  light-front Hamiltonian is
\begin{eqnarray}
 P^-  & = & \sum_{ x^1} \int dx^- 
         - {G^2 \over 4} \Tr\left\{  J^{+}({ x^1})  
        \frac{1}{\partial_{-}^{2}} J^{+}({ x^1}) \right\}
         + {G^2 \over 4N} \Tr \  J^{+}({ x^1}) \frac{1}{\partial_{-}^{2}}
          \Tr \ J^{+}({ x^1}) \; ,
        + V_{ x^1}
\label{lfham} 
\end{eqnarray}
\begin{eqnarray}
J^{+}({x^1}) &=&  i  \left(
M ({ x^1}) \stackrel{\leftrightarrow}{\partial}_{-} 
M^{\da}({x^1})  + M^{\da}({ x^1} - a) 
\stackrel{\leftrightarrow}{\partial}_{-} M({x^1} - a)
\right) \; .
\end{eqnarray}
This is the most general Hamiltonian to order $M^4$ that obeys the
other stated approximations. It can be light-front
quantised and studied in a suitable Fock space at general momenta
$P^+$
and ${P^1}$, as detailed in Ref.~\cite{us2}.
The eigenvalues of the exact Yang-Mills Hamiltonian
yield the glueball masses $M$ through the
relativistic dispersion relation $P^- = (M^2 + ({P^1})^2)/2P^+$.

\subsection{Heavy Sources}

We introduce heavy sources $\phi(x^+,x^-,{x^1})$ on transverse sites. They
are in the fundamental representation and of mass $\rho$.
We apply the same approximations that were made in the pure-glue sector, 
but here we expand to order $M^2$ all operators containing
heavy-source fields, and work at leading non-trivial order in $1/\rho$.
The heavy-source action is $A_s = A + A_{\phi}$ where
\begin{eqnarray}
A_{\phi}  & = & \int dx^+ dx^- \sum_{ x^1}
    \left(D_\alpha \phi\right)^\da D^\alpha \phi 
        - \rho^2 \phi^\da \phi 
- \frac{\tau_1}{N G^2}
\Tr\left\{ F^{\alpha \beta}({x^1}) F_{\alpha \beta} ({x^1})W({x^1}) 
\right\} \nonumber \\
 & & - \frac{\tau_2}{N G^2}
\Tr \left\{ M^{\da}({x^1}) F^{\alpha \beta}({ x^1}) M({x^1})
F_{\alpha \beta}({ x^1} +
a ) \right\} \label{qqlag}
\end{eqnarray}
and
\be
W({x^1}) = \left(M^{\da}({x^1}) M({x^1})+M ({x^1})M^{\da}({x^1})
        \right) \ .
\eq
$D_\alpha = \partial_{\alpha} + i A_{\alpha} ({x^1})$ is the 
usual covariant derivative for the plane $\{x^0, x^2 \}$.
After gauge fixing $A_{-}=0$, eliminating $A_{+}$ in powers of
$M$ from its constraint equation, and discarding the higher orders in $M$,
the  Hamiltonian resulting from $A_s$ which satisfies the
approximations is
\begin{eqnarray}
 P^{-}_{s}  &  = &  \int dx^- \sum_{{x^1}}
 {G^2 \over 4} \Tr\left\{ {J^{+}_{\rm tot} \over \partial_{-}}
{J^{+}_{\rm tot} \over \partial_{-}} 
\right\} 
- {G^2 \over 4N} \Tr \ \left\{ {J^{+}_{\rm tot} \over \partial_{-}} \right\}
\Tr \ \left\{ {J^{+}_{\rm tot} \over \partial_{-}} \right\}
+ V_{{ x^1}} \nonumber \\ && 
+ \rho^2 \phi^\da \phi + \frac{\rho \tau}{a N} \phi^\da W \phi
+ {2 \tau_1 \over N} \Tr\left\{ {J^{+} \over \partial_{-}}
{J^{+} \over \partial_{-}} W 
\right\} 
+ { 2 \tau_2 \over N} \Tr\left\{ {J^{+}({ x^1}) \over \partial_{-}}
M({x^1}) {J^{+}({x^1} + a ) \over \partial_{-}}
M^{\da}({x^1})
\right\} 
\end{eqnarray}
with 
\be
J^{+}_{\rm tot}  = J^+ + i \phi \stackrel{\leftrightarrow}{\partial}_{-}
 \phi^\da  \ .
\eq
Like $P^-$, $P^{-}_{s}$ can be studied in a suitable
Fock space. The eigenvalues of $v^+ P^{-}_{s}$, for co-moving heavy sources of
velocity $v^+$, are the usual excitation energies associated
with the heavy-source potential \cite{burk}. If two sources are
separated by $na$ in the transverse direction $x^1$ and by $L$ in the 
longitudinal direction $x^2$, then
a rotationally invariant string tension would imply that, for large
separations,
\be
v^+ P^{-}_{s} \to  \sigma R \; , \quad R= \sqrt{a^2 n^2 + L^2} 
\eq
for the lowest eigenvalue. Demanding this rotational invariance, then
comparing  results at $n=0$ with $L=0$, allows one to determine
$a$ in a dimensionful unit (we use $G^2 N$) independent of $\sigma$.
This fixes
the relative scale  between $x^1$ and $x^2$, which will be needed for
testing covariance.
In practice it is relatively difficult
to calculate the heavy source potential in the purely transverse direction.
Consequently, we measure the string tension in this direction 
by compactifying space and calculating the winding mode spectrum.

\subsection{Poincar\'e Invariance}

We  test Poincar\'e invariance of the theory by making 
measurements on eigenstates of $P^-$ and $P^{-}_{s}$.
It turns out that a rather simple set of tests suffices to obtain
an accurate estimate of the renormalised trajectory.
One of the approximations made in arriving at $P^-$ (\ref{lfham}) is 
transverse locality. Therefore, it make sense to expand
eigenvalues at 
fixed momenta $(P^+ , {P^1})$ in powers of transverse momentum thus
\be
2P^+ P^- = G^2 N \left( \dm^{2}_{0} 
+ \dm_{1}^{2}\, a^2
     ({P^1})^2 +
\dm_{2}^{2}\, a^4 ({P^1})^4 + 
             \cdots \right) \label{latshell}\; .
\eq
Note that $G$ has dimensions of energy, and $G^2 N$ is held finite in the
$N \to \infty$ limit.\footnote{$aG^2 \to g^2$ as $a \to 0$, but
since we
do not approach $a = 0$ we cannot use the continuum gauge coupling $g$.}
$\dm_0, \dm_1, {\dm}_2,
\cdots$ are dimensionless numbers which we calculate when
diagonalising ${P}^-$.
The simplest requirement of covariance is that
\be 
\dm_{1}^{2}\, a^2 G^2 N -1 
 =  0 \label{cee}
\eq
This ensures isotropy of the speed of light. The dimensionless
quantity $a^2 G^2 N$ has already been determined above from the
scale setting procedure via the string tension. Further conditions
come from higher order corrections in ${P^1}$ in (\ref{latshell}). 
In this work
we will use only the condition (\ref{cee}) for the
lowest-mass glueballs, together with conditions of rotational 
invariance in the heavy-source potential, 
to test the space of couplings of $P^-$. If
our reasoning is correct and our approximations valid, 
we should find a well-defined trajectory on which the
conditions (\ref{cee}) are accurately satisfied --- in practice
we introduce a $\chi^2$ test to quantify this.
Moving
along this trajectory should correspond to changing the spacing
$a$. Eventually this would take us to the transverse continuum limit,
but we will be prevented from reaching $a=0$ by the restriction
$\mu^2 > 0$, a necessary condition for quantisation about $M=0$.

\section{Results}

\subsection{$\chi^2$ charts}

It is convenient to form dimensionless versions of the other
couplings
\be
        m^2 = {\mu^2 \over G^2 N}  \; , \;\; 
        \newl_i = {\lambda_i \over a G^2 N}  \; , \; \; t_i = {\tau_i
\over \sqrt{G^2 N}}.
\eq
The basic technique we follow is to search this space, with the $\chi^2$ test,
for an approximation to a renormalised trajectory on which
observables show enhancement of space-time symmetries violated
by the cut-off $a$.
The $\chi^2$ test is made up of variables to test isotropy of the
speed of light in dispersion of low-lying glueballs, rotational
invariance of the string tension, and rotational invariance
of the potential at intermediate source separations.
Since we can expect to do better with some variables than others,
the weights are adjusted until we produce a sharp trajectory in
coupling space where $\chi^2$ is minimized to roughly one per
effective degree of freedom. In fact, altering the weights
typically changes the sharpness of the trajectory and not
its location. The optimum trajectory is tabulated in Table~\ref{table1}.
Full details on our computations are available at \cite{web}.

\begin{figure}
\centering
\BoxedEPSF{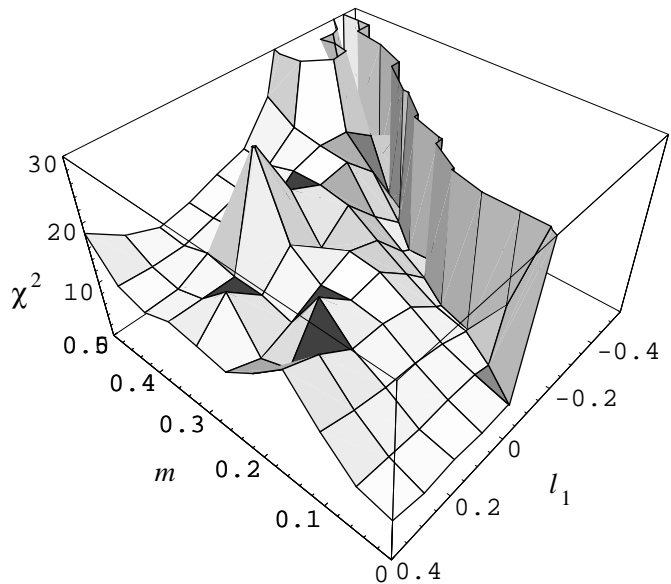 scaled 1000}\\
\caption{Minimum $\chi^2$ for a given $m$ and $\newl_1$. 
In the blank region to the right, tachyons appear in
the spectrum.
\label{fig1}}
\end{figure}

\begin{figure}
\centering
\BoxedEPSF{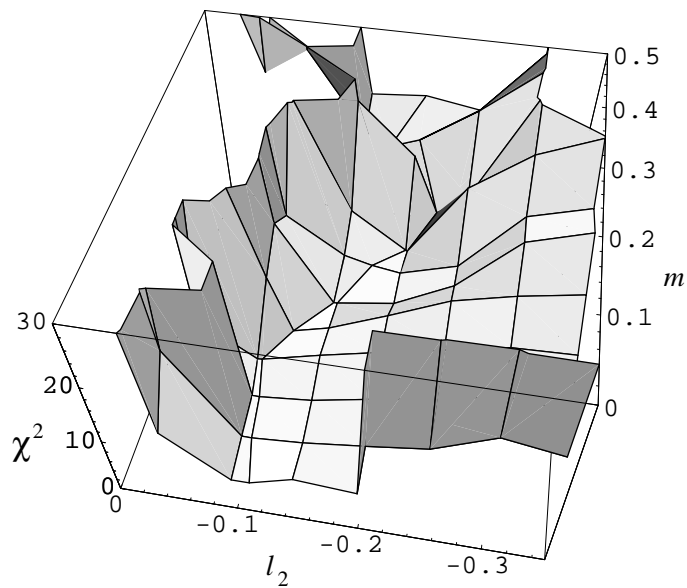 scaled 1000}
\caption{Minimum $\chi^2$ for a given $m$ and $\newl_2$. 
\label{fig2}}
\end{figure}

\begin{figure}
\centering
${a \sqrt{\sigma} }$
\BoxedEPSF{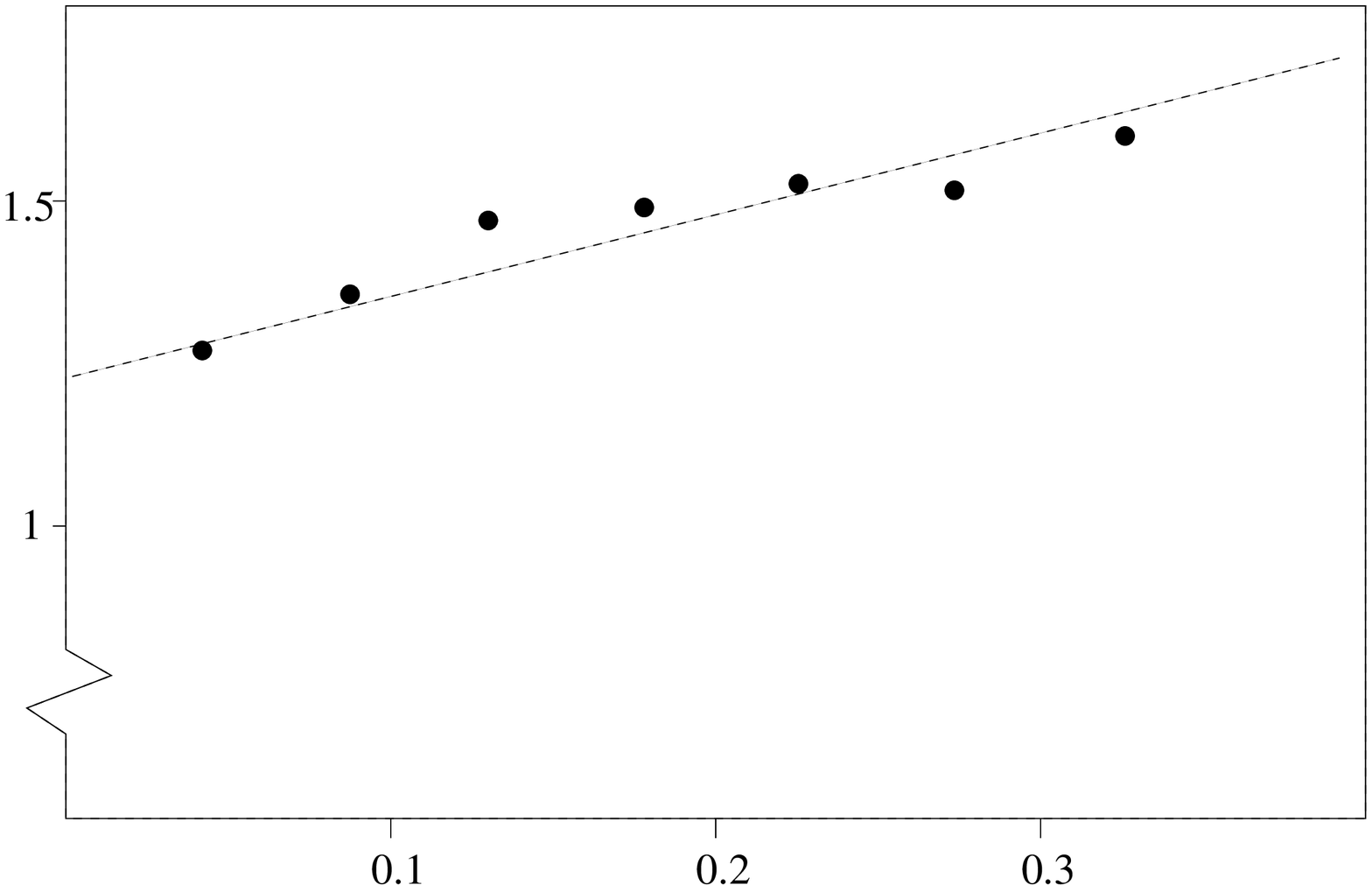 scaled 500}
\\ \hspace{0.5in}$m$
\caption{
Variation of the transverse lattice spacing along the Lorentz
trajectory. The fit is $1.275 m + 1.23$. 
\label{fig3}}
\end{figure}

Figs.~\ref{fig1} and \ref{fig2} show  $\chi^2$ charts for a range of
values of $m$ vs.\ $l_1$ and $m$ vs.\ $l_2$ near the renormalised 
trajectory.\footnote{The behaviour of $l_3$, which is always very large, is
clarified in Ref.~\cite{us2}.}
In each case the renormalised  trajectory appears at the bottom
of a well-defined and  unique $\chi^2$-valley, running from large
to small $m$.
The behaviour of the lattice spacing as one moves along the
renormalised
trajectory is shown in Fig.~\ref{fig3}. As expected, the lattice spacing
gradually decreases with $m^2$ but never becomes zero for $m^2 >0$. 
The fluctuations are due mainly to the difficulty in establishing the
scale $\sqrt{\sigma}$. Since the $\chi^2$ is stable and small over a
range
of lattice spacings, we will use 
the point of smallest lattice spacing ($m=0.044$)
to extract physical quantities.

\subsection{Rotational invariance}

\begin{figure}
\centering
$\displaystyle{v^+ P^- \over \sqrt{G^2 N}}$
\BoxedEPSF{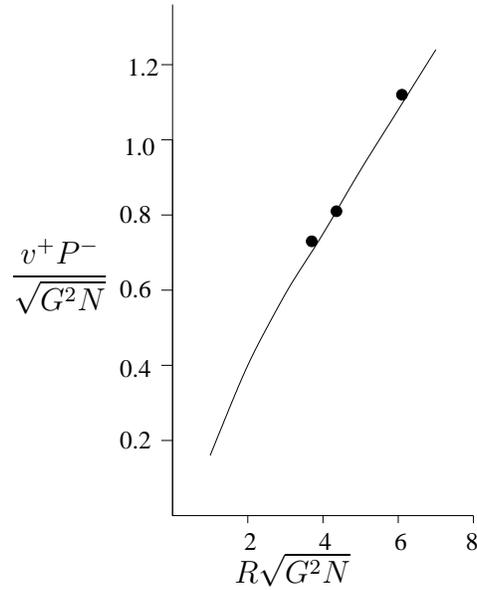 scaled 500}
\\ \hspace{0.5in}$R \sqrt{G^2 N}$
\caption{
The heavy-source potential. Solid line is fit to potential for
sources with $x^2$-separation only; data points are values at one-link
transverse separation and $x^2$-separation $L \sqrt{G^2 N} = 0, 2.5, 5$.
\label{fig4}}
\end{figure}

The heavy-source potential is displayed in Fig.~\ref{fig4}. It shows
better restoration of spatial symmetry than previously 
obtained \cite{us2}. The potential in the 
continuum spatial direction $x^2$ is a fit to the lowest eigenvalue as
a function of $L$ of the form
\be
v^+ P^{-}_{s}   = 0.154\, L G^2 N + 0.183\,\sqrt{G^2 N} - {0.178 \over L}
\; .
\label{care}
\eq
One must be careful when interpreting (\ref{care}) 
since the Coulomb potential in $2+1$
dimensions
is logarithmic. The form (\ref{care}) 
should be appropriate except at the very smallest
$L$, where Coulomb corrections are expected. The $1/L$ term is
a universal correction expected on the grounds of models of flux-string
oscillations \cite{flux}. Universality implies that its coefficient  
should be invariant along the renormalised trajectory. In
fact, we find that it varies slowly, a symptom that our approximation
to the renormalised trajectory is not an exact scaling trajectory for
this quantity and/or
the form (\ref{care}) is not sufficient to fit the potential.

\subsection{Glueballs}
\label{glue}

The spectrum of glueballs in $2+1$ dimensions 
can be classified by $|{\cal J}|^{\cal PC}$, where $\cal J$ is $SO(2)$ spin, 
$\cal C$ is charge
conjugation, and the parity $\cal P$ is spatial reflection $x^1 \to -x^1$. 
Combinations of $\pm {\cal J}$ form parity doublets 
if states are Lorentz covariant. 
On the transverse lattice,
there is enough symmetry to determine $\cal C$, $\cal P$ and 
$|{\cal J}| \bmod{2}$.  Additionally we can examine the shape
of wavefunctions to help distinguish the spin of states. 

\begin{figure}
\centering
$\displaystyle{{\cal M} \over \sqrt{\sigma}}$
\BoxedEPSF{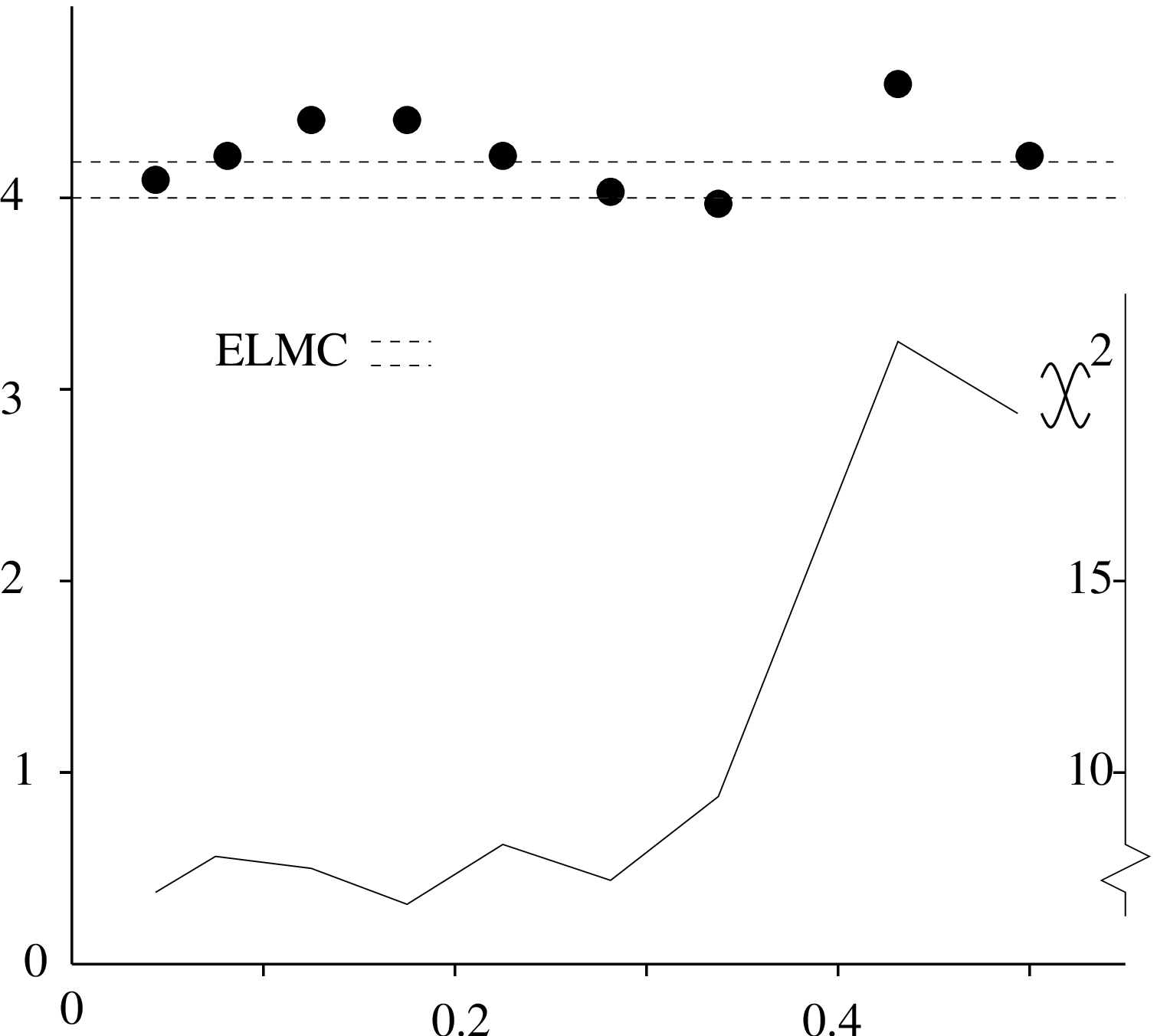 scaled 700}
\\ \hspace{0.5in}$m$
\caption{The variation of the lightest glueball mass along the
renormalised trajectory (together with the associated variation of the
$\chi^2$). Also shown is Teper's extrapolation to $N=\infty$ (ELMC).
\label{fig5}}
\end{figure}

The lightest glueball is a $0^{++}$; its mass along
the renormalised trajectory is shown in Fig.~\ref{fig5}. 
The anisotropy of the speed of light in the $0^{++}$ dispersion 
is less than 3\% for all
the low $\chi^2$ points. For the point of smallest $a$,
$ {\cal M}_{{0}^{++}}= 4.10(13) \sqrt{\sigma}$. Here, we have estimated
a 2--3\% error from extrapolations in DLCQ and
Tamm-Dancoff cut-offs based on known analytic behaviour \cite{us2}, and another
2--3\% from systematic finite-$a$ errors. The fractional
finite-$a$ error estimate is based upon deviations from 
the relativistic dispersion condition (\ref{cee}).
Figure \ref{fig5} also shows the result of Teper, 
${\cal M}_{{0}^{++}} (N \to \infty) = 
4.065(55) \sqrt{\sigma}$, who
fit his finite-$N$ data to a form $A + B/N^{2}$ in order to estimate the
large-$N$ limit. Teper's large-$N$
extrapolation and our independent direct calculation are in
agreement.\footnote{
We note that there are other finite-$N$ lattice results which do not 
agree with Teper's.
Recent Hamiltonian lattice calculations \cite{ham2} yield
${\cal M}_{{0}^{++}} = 3.88(11) \sqrt{\sigma}$  for SU(3)
compared to
${\cal M}_{{0}^{++}} = 4.329(41) \sqrt{\sigma}$ in Ref.~\protect\cite{teper1}.}

\begin{figure}
\centering
\BoxedEPSF{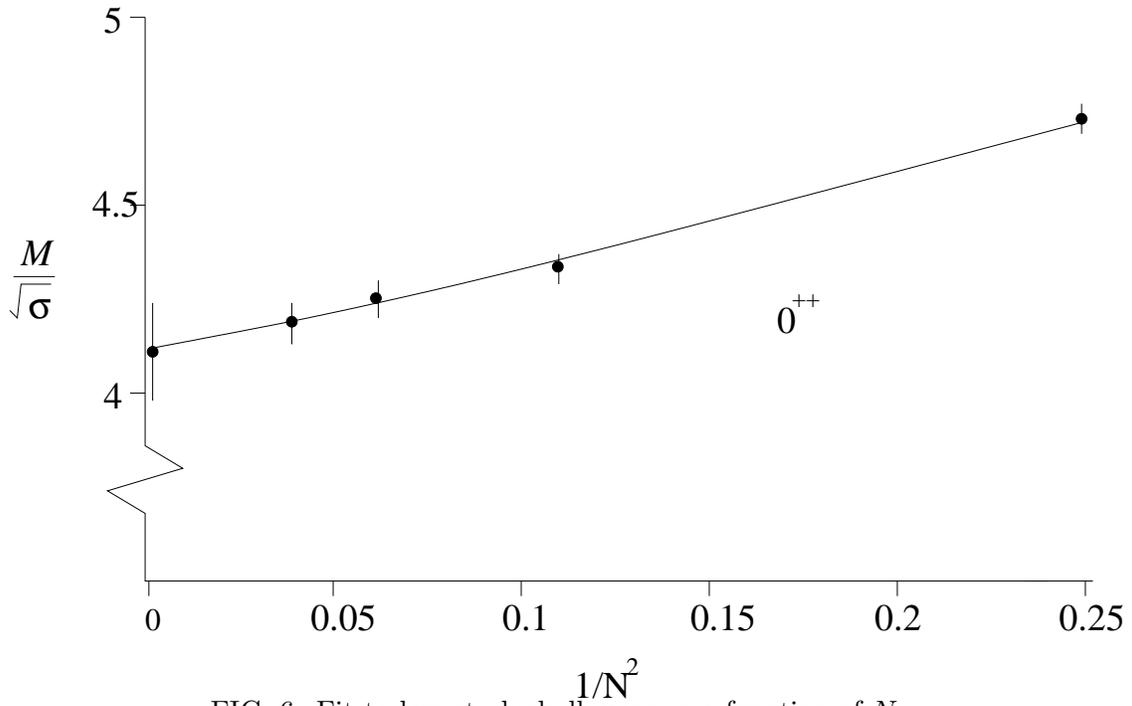 scaled 500}
\caption{Fit to lowest glueball mass as a function of $N$.
\label{fig6}}
\end{figure}

Fitting Teper's finite-$N$ data together with our large-$N$
result we find
\be
{{\cal M}_{{0}^{++}}\over  \sqrt{\sigma}}
= 4.118(13) + {1.55(22) \over N^{2}} + {3.38(73) \over
N^{4}}  \label{ext}
\eq
See Fig.~\ref{fig6}.
This gives a better estimate of the large-$N$ limit than using one or 
the other data set alone.

\begin{figure}
\centering
\BoxedEPSF{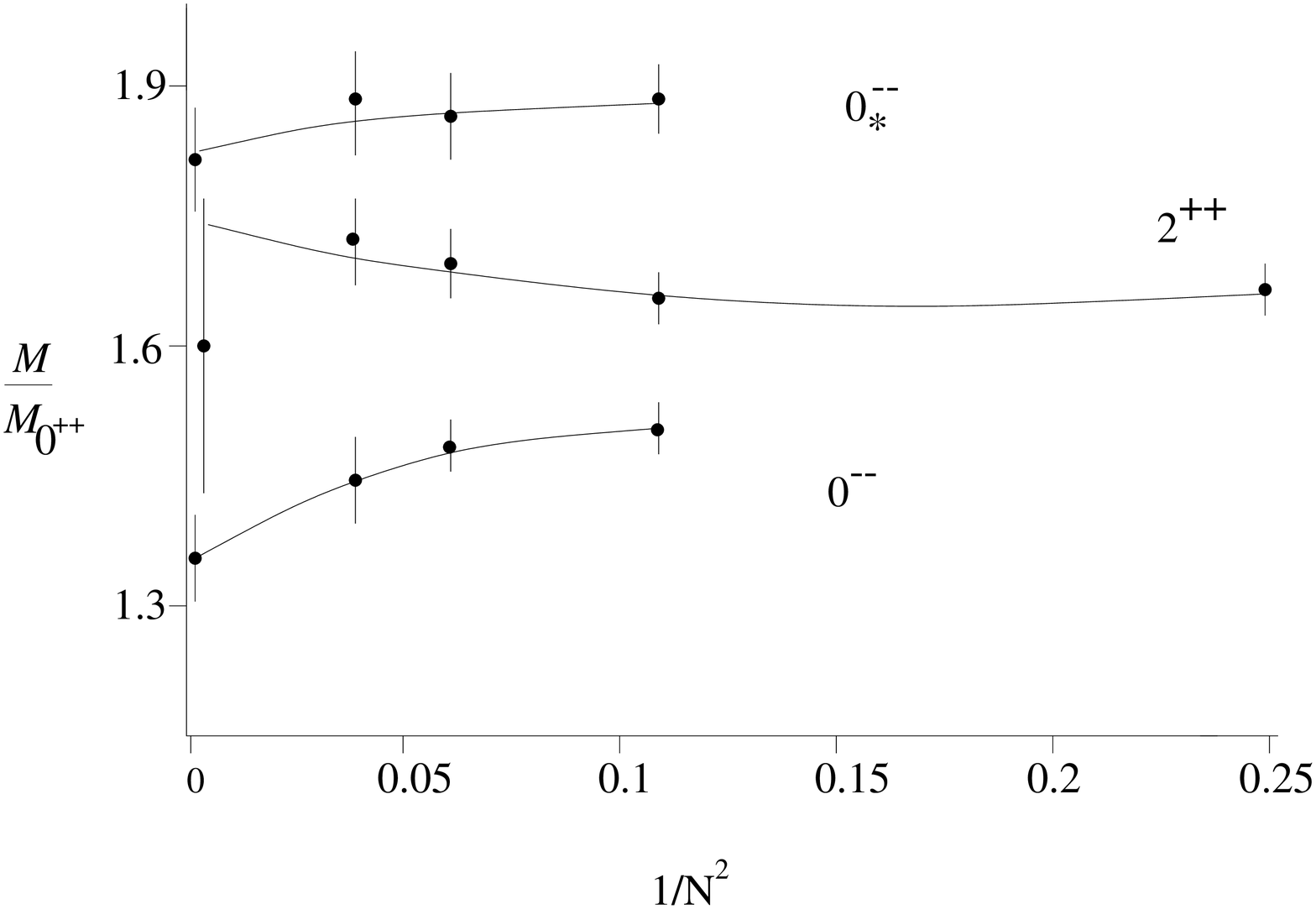 scaled 500}
\caption{Variation of excited  glueball mass ratios with $N$. 
\label{fig7}}
\end{figure}

Although we have made improvements to the
calculation of $\sigma$, the dominant error in Fig.~\ref{fig5}
still comes from the
fluctuation of this quantity; in particular, the determination
of $\sigma$ from the 
longitudinal direction $x^2$ is a big source of error in
determining the relative scales. 
This error becomes so severe for most heavier glueball
states, which exhibit poor
covariance, that an alternative
method must be used for accurate results.
To remove most of the error when dealing
with heavier glueballs, we set ${\cal M}_{0^{++}}/\sqrt{\sigma}$  to the
large-$N$ value estimated in Eqn.~(\ref{ext}), then recalculated the 
renormalised
trajectory with this constraint, {\em id est} we calculate mass
ratios. To improve covariance in the lighter glueballs, at the
expense of heavier states, we also restricted the $\chi^2$ to test only
the dispersion of the lowest states in each charge-conjugation sector.
The resulting mass ratios, at the point of lowest $\chi^2$ on the 
new renormalised trajectory,
are shown in Table \ref{table2}. 
We  also show the fit to the
form  
\be
{{\cal M}\over {\cal M}_{0^{++}}} = A +
\frac{B}{N^{2}} + \frac{C}{N^4}
\eq
including Teper's data. 
The convergence in $1/N^{2}$ is illustrated in Fig.~\ref{fig7}.

\section{Conclusions}

We have found that our improved transverse lattice calculations 
for $2+1$ Yang-Mills in the
large-$N$ limit are consistent with existing finite-$N$ data from an
independent lattice
method. Although both make use of lattice regulators, 
the methods use different quantisation procedures,
elementary degrees of freedom, regulators, gauge fixing, and renormalisation
techniques.
By combining the finite-$N$ and large-$N$ results, we have
obtained accurate estimates of the lightest 
glueball masses and mass ratios for any $N$ in $2+1$ dimensions. 
These should provide a useful 
benchmark for analytic studies. We are able to confirm 
that $O(1/N^{2})$ corrections to the large-$N$ limit are
typically small. The plots of mass versus $1/N^{2}$ could have had any shape,
but they turn out to be almost straight and almost flat.
The same
result also seems to be true in $3+1$ dimensions \cite{us3}, though 
the data is less precise there. Recalling how the
quark model explains the (OZI) suppression of $1/N$ corrections
in most channels \cite{ozi}, it would be interesting to know if
constituent
gluon models could provide an intuitive explanation of our finding.

The lightest glueball mass in units of the
string tension has recently been calculated analytically in
$2+1$ Yang-Mills for any $N$ \cite{nair}. Various
attempts have also been made to 
obtain excited glueball mass ratios 
via extensions the ADS/CFT correspondence at large-$N$ \cite{gravity}. 
These  both use strong-coupling approximations and give qualitatively
reasonable results. The transverse
lattice method is also a strong-coupling one in a  sense;
it is a coarse lattice method. How can these methods give good
results when the large-$N$ approximation almost certainly introduces
a phase transition in the strong-coupling/coarse-lattice regime
\cite{witten}? The answer, we believe, lies in the fact that an exact
renormalised
trajectory, existing in an infinite-dimensional space of couplings, can
avoid such transitions. Thus, it is only necessary to approximate
the renormalised trajectory with a finite number of couplings to
obtain results relevant to the continuum limit. One cannot extrapolate
to weak coupling or fine lattice limits, but if one 
chooses the right variables and couplings, results can be obtained directly
from the approximation to the renormalised trajectory far from these
limits.

\vspace{5mm}

{\bf Acknowledments} SD is supported by PPARC grant 
No.\ GR/LO3965.  BvdS was supported by an award from Research
Corporation.
Computations were performed at the 
Ohio Supercomputer Center and at the Pittsburgh Supercomputer
Center.

\newpage 

\begin{table}
\centering$\displaystyle
\renewcommand{\arraystretch}{1.25}
\begin{array}{|cccccc|c|}
\hline
 m & l_1 & l_2 & l_3 & t_1 & t_2 & \chi^2 
\\[7.5pt]\hline\hline
0.044 & -0.052 & -0.112 & 680.2 & -0.661 & -0.691 & 7.02 \\
0.089 & -0.087 & -0.109 & 396.8 & -0.780 & -0.811 & 7.85\\
0.134 & -0.108 & -0.091 & 3.221 & -0.896 & -0.876 & 7.56\\
0.180 & -0.147 & -0.107 &4.401 & -0.943 & -0.927 & 6.55\\
0.226 &  -0.204& -0.134 & 178.5 &-1.098  & -1.167 &8.02 \\
0.2765 &-0.240  &-0.153  & 5.48  & -0.989& -1.138 & 7.31\\
0.3275 & -0.308 & -0.157 & 6.01& -1.181 & -1.340 & 8.64\\
\hline
\end{array}$
\caption{The trajectory in coupling-constant space 
which minimises the $\chi^2$ test of covariance.
\label{table1}}
\end{table}

\begin{table}
\centering$\displaystyle
\renewcommand{\arraystretch}{1.25}
\begin{array}{|c|c|ccc|}
\hline
 \displaystyle |{\cal J}|^{\cal P C} & {\cal M}/{\cal M}_{0^{++}} & 
               \multicolumn{3}{c|}{\mbox{Fit coefficients}}\\[-1em]
 &  & C & B  & A \\ \hline \hline
0^{--} & 1.35(5)  & -14.58(1.47) & 2.983(191) & 1.349(6) \\ 
2^{++} & 1.60(17)  & 3.233(2.724) & -1.144(856) &  1.743(51)   \\ 
0^{--}_{*}   & 1.82(6) & -5.839(7.488) & 1.136(941) & 1.824(25)   \\
2^{-+}  & 1.77(?) & - & 0.659(246)  & 1.697(57)   \\       
0^{++}_{*} & 1.28(?) & - & 0.770(399) & 1.520(38)   \\       
\hline 
\end{array}$
\caption{Mass ratios for lightest glueball excited states, showing our
$N = \infty$ measurement and fit coefficients including finite-$N$
data from Ref.~\protect\cite{teper1}. The $2^{-+}$ and $0^{++}_{*}$
states were not covariant-enough for reliable error estimates,
and only Teper's finite-$N$ extrapolation is shown.
\label{table2}}
\end{table}

\vfil

\begin{references}


\bibitem{teper1} M. Teper, Phys.\ Rev.\ D {\bf 59}, 014512 (1999).

\bibitem{ham1} C. Hamer, M. Sheppeard, W-H. Zheng, and D. Schutte,
Phys.\ Rev.\ D {\bf 54}, 2395 (1996); \\
C. Hamer,  Phys.\ Rev.\ D {\bf 53}, 7316 (1996). 

\bibitem{ham2}  Q. Z. Chen, X. Q. Luo, S. H. Guo, and X. Y. Fang, 
Phys.\ Lett.\ B {\bf 348}, 560 (1995);\\
 X. Q. Luo, Q. Z. Chen, S. H. Guo, X. Y. Fang, J. Liu, 
Nucl.\ Phys.\ B {\bf 53}, (Proc.\ Suppl.) 239 (1997).

\bibitem{nair} D. Karabali and V. P. Nair, Phys.\ Lett.\ B {\bf 379},
141 (1996); Nucl.Phys.\ B {\bf 464}, 135 (1996);
Int.\ J. Mod.\ Phys.\ A {\bf 12} 1161 (1997);\\
D. Karabali, C. Kim, and V. P. Nair,
Phys.\ Lett. B {\bf 434},  103 (1998); Nucl.Phys.\ B {\bf 524}, 661 (1998).

\bibitem{gravity} C. Csaki, H. Ooguri, Y. Oz, and J. Terning,
JHEP {\bf 9901},  017 (1999);\\
R. de Mello Koch, A. Jevicki, M. Mihailescu, and J. P. Nunes,
Phys.\ Rev.\ D {\bf 58}, 105009 (1998);\\
R. Brower, S. Mathur, and C-I. Tan, {\tt hep-th/0003115}.


\bibitem{hoof} G. 't Hooft, Nucl.\ Phys.\ B {\bf 72}, 461 (1974).


\bibitem{teper2} M. Teper, Phys.\ Lett.\ B {\bf 397}, 223 (1997); 
           Phys.\ Lett.\ B {\bf 311}, 223 (1993);
Phys.\ Lett.\ B {\bf 289}, 115 (1992).

\bibitem{marek} S. A. Chin and M. Karliner,  Phys.\ Rev.\ Lett.\ {\bf 58},
1803 (1987).

\bibitem{griffin} P. A. Griffin, Nucl.\ Phys.\ B {\bf 139}, 270 (1992).


\bibitem{us1} S. Dalley and B. van de Sande,  Nucl.\ Phys.\ B {\bf 53} 
(Proc.\ Suppl.),  827 (1997);  Phys.\ Rev.\ D {\bf 56},
7917 (1997); 

\bibitem{us2} S. Dalley and B. van de Sande, Phys.\ Rev.\ D {\bf 59},
065008 (1999).


\bibitem{bard}  W. A. Bardeen and R. B. Pearson, 
              Phys.\ Rev.\   D {\bf 14}, 547 (1976);\\
W. A. Bardeen, R. B. Pearson, and E. Rabinovici, 
           Phys.\ Rev.\  D {\bf 21},  1037 (1980). 

\bibitem{us3} S. Dalley and B. van de Sande, Phys.\ Rev.\ Lett.\ {\bf 82},
1088 (1999);  Phys.\ Rev.\  D {\bf 62}  014507 (2000).





\bibitem{burk} M. Burkardt and B. Klindworth,  Phys.\ Rev.\ 
D {\bf 55},  1001 (1997).

\bibitem{dlcq} H.-C.\ Pauli and S. J.  Brodsky, 
          Phys.\ Rev.\ D {\bf 32}, 1993 and 2001 (1985).

\bibitem{perry} R. J. Perry and K. G. Wilson, Nucl. Phys. B {\bf 403},
587 (1993).

\bibitem{web} http://www.geneva.edu/\~{}bvds/dave/

\bibitem{flux} M. Luscher,  Nucl.\ Phys.\  B {\bf 180},
 317 (1981).

\bibitem{ozi} P. Geiger and N. Isgur, Phys.\ Rev.\ D {\bf 44}, 799
(1991);
Phys.\ Rev.\ Lett.\ {\bf 67}, 1066 (1991); Phys.\ Rev.\ D {\bf 47}, 5050
(1993);

\bibitem{witten} D. J. Gross and E. Witten, 
Phys.\ Rev.\ D {\bf 21}, 446 (1980).


\end{references}
\end{document}